%% file: coherence.tex
\documentclass[a4paper,twocolumn,accepted=2021-01-23]{quantumarticle}

\input{config.tex}

\begin{document}

\title{Searching for Coherent States: \newline From Origins to Quantum Gravity}

\author{Pierre Martin-Dussaud}
\email{pmd@cpt.univ-mrs.fr}
\affiliation{Aix Marseille Univ, Université de Toulon, CNRS, CPT, Marseille, France}
\affiliation{Basic Research Community for Physics e.V.}
\orcid{0000-0002-9213-8036}

\maketitle 

\begin{abstract}
\noindent
We discuss the notion of coherent states from three different perspectives: the seminal approach of Schrödinger, the experimental take of quantum optics, and the theoretical developments in quantum gravity. This comparative study tries to emphasise the connections between the approaches, and to offer a coherent short story of the field, so to speak. It may be useful for pedagogical purposes, as well as for specialists of quantum optics and quantum gravity willing to embed their perspective within a wider landscape.
\end{abstract}

\section{Introduction}

Coherent states are essential tools in theoretical physics. Since their early introduction by Schrödinger in 1926, they have served practical purposes in quantum optics, while several mathematical generalisations of the notion have been proposed, and some of them applied to quantum gravity. The present paper was initially motivated by the following observations:
\begin{itemize}
\item The existing reviews of coherent states, like \cite{gazeau2009} or \cite{perelomov1986}, do not deal with quantum gravity. So, we would like to summarise the various coherent states introduced in quantum gravity.
\item The quantum gravity literature is very technical and does not insist much on the conceptual motivations behind the definitions. We would like to show that the semi-classical properties of coherent states are expected rather than magical.
\item The many approaches to coherent states convey the impression of a disparate field made of arbitrary definitions. On the contrary, we would like to insist on the unity of the landscape and expose the big picture.
\end{itemize}
Thus we offer a journey among coherent states, from Schrödinger to quantum gravity, passing by quantum optics, always triggered both by conceptual clarity and concision. The resulting paper has this kind of hybrid format, between the review and the pedagogical introduction, trading exhaustiveness for clarity. It may be of interest for both communities of quantum gravity and quantum optics as it explains to the one what has been done by the other.

Along the way, we will notably answer the following puzzles:
\begin{itemize}
\item Coherent states are usually introduced for the harmonic oscillator, but can't we define them for even simpler systems like the free particle?
\item Coherent states are sometimes presented as the states $\ket{\psi}$ such that $\ev{\hat x(t)}{\psi}$ and $\ev{\hat p(t)}{\psi}$ satisfy the classical equations of motion. It is, for instance, the impression conveyed in the seminal paper of Schrödinger \cite{Schrodinger1926}, but also in the recent reference book \cite{gazeau2009}. However such a property cannot be a characterisation of coherent states whatsoever, since it is clear, from Ehrenfest theorem, that \textit{any} time-evolved state $\ket{\psi(t)}$, coherent or not, satisfies it. Is there a way to make this first intuition of classicality rigorous?
\item Coherent states are also often introduced as eigenstates of the annihilation operator, but this does not seem to be the best pedagogical way as the physical motivation of this approach may seem rather obscure at first sight. Indeed, doing so, the classical properties that can be checked afterwards appear as magical, rather than expected. What could be a better pedagogical introduction to the topic?
\item Coherent states can also be generated by the action of the Heisenberg group over the vacuum state. This group is sometimes called the dynamical symmetry group of the harmonic oscillator (see \cite{gilmore1974, feng1990}), although it is very unclear in which sense the group is "dynamical", a "symmetry group", or even specific to the harmonic oscillator. Can we make the statement precise?
\item Coherent states are wanted to be quasi-classical states, but in quantum optics, for instance, the coherent states of light are those that maximise the interference pattern, which is paradoxically regarded as a very quantum feature, far from being a classical source of light as an incandescent bulb may be. Is the paradox of designation only superficial? 
\item The definition of coherent states in quantum gravity is covered by a jungle of technicalities, far from the experimental point of view of quantum optics. Can we nevertheless summarise the story to keep the key physical idea and make our way through the jungle?
\end{itemize}

To start with, we go back to the initial ideas of Schrödinger in section \ref{sec:schroedinger coherent states}, and propose a modern follow-up in section \ref{sec:dynamical characterisation}. Then, in section \ref{sec:kinematical characterisation}, we enlarge the discussion with a kinematical characterisation of coherent states in terms of annihilation operators. We explain the physical meaning of these operators in quantum optics in section \ref{sec:optical coherence}, which motivates an algebraic generalisation of coherent states presented in section \ref{sec:algebraic approach}. In section \ref{sec:geometric approach}, we present an independent geometric generalisation, which was later applied in quantum gravity, as we show in section \ref{sec:quantum gravity}.

\section{Schrödinger coherent states} \label{sec:schroedinger coherent states}

Historically, the initial motivation for introducing coherent states is to demonstrate how classical mechanics can be recovered from quantum mechanics. It is done in 1926 in a short seminal paper by Erwin Schrödinger \cite{Schrodinger1926}, translated in English in \cite{Schrodinger1928}, entitled \textit{The Continuous Transition from Micro- to Macro-Mechanics}. The title is rather explicit about its goal, although one may discuss whether it has been achieved or not.

Interestingly, Schrödinger does not use the word "coherent" anywhere, but he aims at constructing mathematically
\begin{quote}
\textit{a group of proper vibrations [that] may represent a "particle", which is executing the "motion", expected from the usual mechanics}.
\end{quote}
Neither does he use the words "quasi-classical" or "semi-classical", but the latter would convey his intuition probably better than "coherent". The paper does not shine by its clarity, but one can understand the overall logic, that we present below in modernised terms and notations. 

\passage{Quantum harmonic oscillator}

Let's consider the quantum harmonic oscillator in one dimension, with mass $m$ and pulsation $\omega$. Its Hilbert space is $L^2(\mathbb{R})$ over which are acting the position operator $\hat x \psi(x) = x \psi(x) $ and the momentum $\hat p \psi  = - i \hbar \, \partial_x \psi $ . The dynamics is provided by the hamiltonian which reads 
\begin{equation}
\hat H = \frac{\hat p^2}{2m} + \frac{1}{2} m \omega^2 \hat x^2.
\end{equation} 
The eigenstates of $\hat H$ form an orthonormal basis $\ket{n}$, indexed by $n \in \mathbb{N}$, whose coefficients, in the basis of eigenstates $\ket{x}$ of $\hat x$, read
\begin{equation}
\braket{x}{n} = \sqrt[4]{\frac{m \omega}{\pi \hbar}} \frac{1}{\sqrt{2^n n!}} e^{- \frac{m \omega}{2 \hbar} x^2} H_n \left(\sqrt{\frac{m \omega}{\hbar}} x \right)
\end{equation}
where $H_n(x)$ are Hermite's polynomials\footnote{Wikipedia mentions two conventions for Hermite's polynomial. We use the physicist one, i.e.
\begin{equation}
H_n(x) \overset{\rm def}= (-1)^n e^{x^2} \dv[n]{}{x} e^{-x^2}.
\end{equation}} and the associate eigenvalues are 
\begin{equation}
E_n = \hbar \omega \left( n + \frac{1}{2} \right).
\end{equation}

\passage{Schrödinger coherent states}

Then, Schrödinger defines\footnote{Compared to the strictly original definition of Schrödinger, we have here chosen to normalise the states, with a factor $e^{- A^2/2 } $ in front of the sum, and a phase factor $e^{- i \omega t /2}$ to match the standard Dirac notation.}, out of the blue, the following family of states, indexed by time $t \in \mathbb{R}$ and another parameter $A \in \mathbb{R}$:
\begin{equation}\label{eq:schrodinger CS}
 \ket{A e^{i \omega t}} \overset{\rm def}= e^{- \frac{A^2}{2}} e^{- i \omega t /2} \sum_{n=0}^\infty \frac{A^n}{\sqrt{n!}}  e^{ \frac{i}{\hbar} E_n t}  \ket{n}.
\end{equation}
It is immediate to see that $\ket{A e^{i \omega t}}$ is the temporal evolution of $\ket{A}$ by the unitary operator $e^{\frac{i }{\hbar} \hat H t}$, as
\begin{equation}
e^{\frac{i}{\hbar} \hat H t} \ket{A} = e^{i \omega t /2} \ket{A e^{i \omega t}}.
\end{equation}
Then Schrödinger argues that these states approximate the "macro-mechanics", what we would call in modern language, being semi- or quasi-classical. More precisely, he highlights three properties:
\begin{enumerate}
\item First the average position satisfies the law of classical motion:
\begin{equation}\label{eq:classical motion}
\ev{\hat x} = A \cos \omega t.
\end{equation}
\item Secondly, the average energy is almost the classical one:
\begin{equation}
\ev{\hat H} \approx  A^2  m \omega^2.
\end{equation}
\item Third, he argues (without any explicit computation) that the wave packet does not "spread out", but "remains compact", like a particle. 
\end{enumerate}

The two first properties provide physical meaning to the parameter $\alpha$ as the amplitude of some corresponding classical wave. Thus the Schrödinger coherent states are parametrised by an amplitude $\alpha$ and an instant $t$.

\passage{Wrong characterisation of quasi-classicality}

The three arguments above appear as a first attempt to formalise the property of "quasi-classicality", and have been the basis of the later developments of coherent states. Unfortunately, it has been hardly never noticed that the first property cannot characterise quasi-classicality in any way. Indeed \textit{all} the quantum states of the harmonic oscillator satisfy this property! More precisely, given any initial state $\ket{\psi_0}$, its time evolution will be so that it satisfies equation \eqref{eq:classical motion}. It is a consequence of Ehrenfest theorem, that drives the evolution of the expected value of a time-independent observable $\hat A$ in any state $\ket{\psi(t)}$, according to the equation
\begin{equation}
\dv{\ev*{\hat A}}{t} = \frac{1}{i\hbar}  \ev{[\hat{A},\hat{H}]}.
\end{equation}
In the case of the harmonic oscillator, the equations for $\hat x$ and $ \hat p$ are
\begin{align}
\dv{\ev*{\hat x}}{t} = \frac{1}{m} \ev{\hat p} && \text{and} &&
\dv{\ev*{\hat p}}{t} = - m \omega^2 \ev{\hat x}.
\end{align}
These are actually the classical equations of motion for $\ev{\hat x}$ and $\ev{ \hat p}$, and so all solutions $\ev{\hat x}$ take the form of equation \eqref{eq:classical motion}. It is completely generic and so cannot be used as a characterisation of quasi-classicality\footnote{I am indebted to Federico Zalamea to have made me realised this fact.}. Thus, there is nothing such as a constraining property in Schrödinger's first statement, except maybe the implicit demand that the time evolution of a "quasi-classical state" should still be "quasi-classical". It is surprising that this fact has not been much recognised, and that many recent developments of coherent states still treat this property as an argument for the "peakiness" of the coherent states. Of course, with a more complicated hamiltonian, the property is not a trivial statement, but for the harmonic oscillator, it is.

Let us now analyse the two other properties, $3$ and $2$, which at first may disappoint us with their vague formulation. When they are made precise, we show that each of them, alone, is a sufficient condition that fully characterises the family of coherent states.  

\section{Dynamical characterisation} \label{sec:dynamical characterisation}

A characteristic feature of quantum mechanics is the fact that the position of a particle is not given by a classical trajectory, but rather by a probability density that evolves with time. Thus, a "quasi-classical" state could be one for which a quantum particle is \textit{well localised in space}, and \textit{remains localised as time goes by}. Let us try to formalise it, and see how this programme fails in the case of the free particle and succeeds for the harmonic oscillator.

\passage{Free particle}

Consider the free particle in one dimension. Its Hilbert space is $L^2(\mathbb{R})$. The Dirac delta function $\delta(x)$ describes the state of a particle perfectly well localised at $x=0$. The uncertainty about its position is zero: $\Delta \hat x = 0$. For that reason, it may seem a good candidate for being a quasi-classical state.

However, this first attempt fails because the particle does not remain localised as time goes by. Indeed, the hamiltonian of the free particle
\begin{equation}
\hat H = \frac{\hat p^2}{2m},
\end{equation}
drives the time-evolution of $\delta(x)$ to
\begin{equation}\label{eq:free particle delta}
\psi(x,t) = \sqrt{\frac{m}{2\pi \hbar |t|}} e^{-i \, {\rm sgn} (t) \,  \frac{\pi}{4}} e^{i \frac{m x^2 }{2 \hbar t}}.
\end{equation}
The probability distribution $|\psi(x,t)|^2$ is now completely spread, $\Delta \hat x = \infty$, and not even normalised! Thus, a wave function which is infinitely well localised at initial time, turns instantaneously into an infinitely spread state\footnote{This matter of fact seems even to contradict the postulate according to which two successive measurements should give the same result. But these pathologies can be imputed to the already suspicious Dirac delta function.}.

So consider instead a more reasonable initial state, like a gaussian curve 
\begin{equation}
\psi_0(x) = \frac{1}{\sqrt{\sigma \sqrt{2 \pi}}} e^{- \frac{x^2}{4 \sigma^2}},
\end{equation} 
which is spread as $\Delta_0 \hat x = \sigma$. It evolves as a free particle to
\begin{equation}
\psi(x,t) = \frac{1}{ \left( 2\pi (\sigma^2 + i \hbar t /m ) \right)^{1/4}} e^{- \frac{x^2}{4 (\sigma^2 + i \hbar t /m)}}.
\end{equation}
It is also a gaussian which is spread like  
\begin{equation}
\Delta \hat x  = \sqrt{\sigma^2 + \frac{t^2 \hbar^2}{4 m^2 \sigma^2}},
\end{equation}
so that it irrevocably spreads with time and looses its initially compact shape.

In fact, whatever the initial state $\psi_0$ at time $t=0$, it evolves as a free particle to a state $\psi(x,t)$ which satisfies\footnote{A proof can be found in \cite{hall2013} p. 104.}:
\begin{multline}
(\Delta \hat x)^2 =  (\Delta_0 \hat x )^2 + \frac{(\Delta_0 \hat p)^2}{m^2} t^2 \\ + \frac{t}{m} \left(  \ev{\hat x \hat p + \hat p \hat x}_0 - 2 \ev{\hat x}_0 \ev{\hat p}_0  \right) .
\end{multline}
It is a second order polynomial in $t$. A necessary condition to prevent the time spreading would be to have $(\Delta_0 \hat p)^2=0$, but this implies, through Heisenberg inequality, that $\Delta_0 \hat x = \infty$, i.e. a maximally spread states in space... So, for the free particle, the spreading is unavoidable. From this perspective, there is no "quasi-classical state" for the free particle. 

\passage{Harmonic oscillator}

Let us now consider the more sophisticated hamiltonian of the harmonic oscillator:
\begin{equation}
\hat H = \frac{\hat p^2}{2m} + \frac{1}{2} m \omega^2 \hat x^2.
\end{equation}
A priori, there is more chance to find coherent states now, because we have added a potential well in the hamiltonian that can help to confine the wave-function and prevent it from spreading.
In this case, the spreading of a general solution $\psi(x,t)$ is given by\footnote{\begin{proof}
Using Ehrenfest theorem we have
\begin{equation}
\begin{array}{l}
\frac{d}{dt} \ev{\hat x} =\ev{\hat p}/m \\
\frac{d}{dt} \ev{\hat p} = - m \omega^2 \ev{\hat x}\\
\frac{d}{dt} \ev{\hat x^2}= \ev{\hat x\hat p+\hat p\hat x}/m\\
\frac{d}{dt} \ev{\hat p^2} = - m \omega^2 \ev{\hat x\hat p+\hat p\hat x} \\
\frac{d}{dt} \ev{\hat x\hat p + \hat p\hat x} = 2 \ev{\hat p^2} /m - 2 m \omega^2 \ev{\hat x^2}
\end{array}
\end{equation}
from which we show the differential equation
\begin{equation}
\frac{d^4}{dt^4} \ev{\hat x^2} = - 4 \omega^2 \frac{d^2}{dt^2} \ev{\hat x^2}
\end{equation}
which is finally solved easily, and leads to our expression for  $(\Delta \hat x)^2 \overset{\rm def}=  \ev{\hat x^2} - \ev{\hat x}^2  $.
\end{proof}}
\begin{multline}
(\Delta \hat x)^2 = \frac{1}{2} \left( (\Delta_0 \hat x)^2 + \frac{(\Delta_0 \hat p)^2}{m^2 \omega^2} \right) \\
- \frac{1}{2} \left(\frac{(\Delta_0 \hat p)^2}{m^2 \omega^2} -  (\Delta_0 \hat x)^2  \right) \cos (2 \omega t) \\
+ \frac{1}{2m\omega} \left(  \ev{\hat x\hat p + \hat p\hat x}_0 - 2 \ev{\hat x}_0 \ev{\hat p}_0  \right) \sin (2 \omega t) 
\end{multline}
It is noticeable that the spreading oscillates. There is no irresistible increasing of the spreading. Instead, whatever the state we start with, the wave packet will stay confined within a finite range, and even come back periodically to its initial spreading $\Delta_0\hat x$. 

So a first lesson to draw from this computation is that one should not be surprised by the fact that Schrödinger coherent states do not spread out, as a free particle would do, because no state of the harmonic oscillator does it! 

\passage{Constant and minimal}

Then, one can try to express the third property of Schrödinger in more precise terms. For instance, one can look for states for which $\Delta \hat x$ is constant in time. This requires the two conditions
\begin{equation}\label{eq:condition constant}
\left\{
\begin{array}{l}
(\Delta_0 \hat p)^2 = m^2 \omega^2 (\Delta_0 \hat x)^2 \\
\ev{\hat x\hat p + \hat p\hat x}_0 = 2 \ev{\hat x}_0 \ev{\hat p}_0
\end{array}
\right.
\end{equation}
One can check that the Schrödinger coherent states do satisfy these conditions. However, these two conditions are not sufficient to characterise them. For instance all the eigenstates of the hamiltonian, $\ket{n}$, also satisfy these conditions. Another condition is still required: the minimisation of $\Delta_0 \hat x$.

The family of states for which $\Delta \hat x$ is constant in time is foliated by the value of $\Delta_0 \hat x$, with a minimal value being strictly positive. Indeed, from the Heisenberg inequality
\begin{equation}
\Delta_0 \hat x \Delta_0 \hat p \geq \frac{\hbar}{2}.
\end{equation}
and from the first condition in \eqref{eq:condition constant}, we have
\begin{equation}
\Delta_0 \hat x  \geq \sqrt{\frac{\hbar}{2m \omega}}.
\end{equation}
Now one can show that the only states minimising this inequality are the coherent states of Schrödinger! We have thus found a characterisation of them: they are these states whose spreading in position $\Delta \hat x$ is \textit{constant and minimal}. Both conditions are important. Otherwise, there are states whose spreading is momentarily smaller but will grow later to a larger value. There are also states whose spreading is constant, but not minimal (like the $\ket{n}$). Geometrically, the two conditions select a $2$-dimensional submanifold out of the infinite-dimensional space $L^2(\mathbb{R})$.

\passage{Minimal time average}

There is another way to make the third property of Schrödinger more precise. Consider the time average of $\Delta \hat x$:
\begin{equation}
T \left[ \Delta \hat x  \right] = \sqrt{ \frac{(\Delta_0 \hat x)^2}{2}  + \frac{(\Delta_0 \hat p)^2}{2 m^2 \omega^2} }.
\end{equation}
Now, from Heisenberg inequality, this time average is bounded by
\begin{equation}
T\left[ \Delta \hat x  \right] \geq \sqrt{\frac{\hbar}{2m \omega}}.
\end{equation}
This inequality is saturated for the coherent states and only for them. So we have a second characterisation of coherent states as the states which minimise $\Delta \hat x$ on average (as time goes by).

\passage{"Almost classical energy"}

Let us now turn to the second property underlined by Schrödinger: "the average energy is almost classical".
As we said before, the classical behaviour of $\ev{\hat x}$ cannot reasonably be taken as  evidence for the classicality of coherent states. So one may wonder whether the same holds for $\ev*{\hat H}$. It is easy to see that
\begin{multline}
\ev{\hat H(\hat x, \hat p) } =  H(\ev{\hat x}, \ev{\hat p}) \\ + \frac{1}{2m} (\Delta \hat p)^2 + \frac{1}{2} m \omega^2 (\Delta \hat x)^2,
\end{multline}
where
\begin{equation}
H(x,p) \overset{\rm def}= \frac{1}{2m}p^2 + \frac{m \omega^2}{2} x^2
\end{equation}
is the classical hamiltonian, function over the phase space $\mathbb{R}^2$ with coordinates $(x,p)$.
From Heisenberg inequality one deduces then that
\begin{equation}
\ev{\hat H(\hat x, \hat p) } -  H(\ev{\hat x}, \ev{\hat p}) \geq \frac{\hbar \omega}{2}.
\end{equation}
One can say in precise terms that a state is quasi-classical with respect to the energy if it saturates this inequality. 
And happily, this condition alone is sufficient to define the coherent states!

\passage{Conclusion}

We have cleaned up the properties of coherent states underlined by Schrödinger in his seminal paper. First, we have realised that the first property was very generic and not specific to coherent states. Secondly, a careful analysis of the two other properties has led us to formulate three equivalent definitions of coherent states of the harmonic oscillator:
\begin{enumerate}
\item Constant and minimal $\Delta \hat x$
\item Minimal temporal average of $\Delta \hat x$
\item Minimal $\left(\ev*{\hat H(\hat x, \hat p) } -  H(\ev{\hat x}, \ev{\hat p})\right)$.
\end{enumerate}
We regard these definitions as better suited for a pedagogical introduction to coherent states, compared to the abstract definition as eigenstates of the annihilation operator that one finds in most textbooks.  

These three definitions are \textit{dynamical} in the sense that they make use of the temporal evolution of states or the hamiltonian. A priori, if another hamiltonian is used, like for the free particle, another family of states will be found. In this sense, one can talk indeed of the coherent states \textit{of the harmonic oscillator}, and not \textit{of the free particle}. The stability of the family of coherent states under temporal evolution is made obvious with the two first definitions, but not so much with the third one.

In an attempt of generalisation of the notion of coherent states, we are now going to relax this dynamical aspect of the definitions and propose a purely kinematical characterisation.

\section{Kinematical characterisation} \label{sec:kinematical characterisation}

Let's start all over again, from a general quantum system. Its states form a Hilbert space $\mathcal{H}$, and we consider the problem of finding the states which are "quasi-classical" in a sense to be determined.

\passage{Geometrical formulation}

As regards its kinematical features, the departure of quantum mechanics from classical mechanics can be understood geometrically, through the so-called \textit{geometrical formulation of quantum mechanics} \cite{Schilling1996, Zalamea2018}.

In quantum mechanics, we are used to systems whose states are taken to be vectors of a Hilbert space $\mathcal{H}$ endowed with a scalar product $\braket{.}{.}$, and the algebra of observables $\mathcal{B}_\mathbb{R}(\mathcal{H})$ consists of the (bounded) self-adjoint linear operators over $\mathcal{H}$. In fact, we only consider the normalised vectors of $\mathcal{H}$, up to a global phase, so that the space of physical states really is the projective Hilbert space $P \mathcal{H}$. The geometric features of this space enable a fair comparison to the classical phase space. $P \mathcal{H}$ is a Khäler manifold which means that it is naturally endowed with two geometric structures: a symplectic $2$-form $\omega$ (coming from the imaginary part of $\braket{.}{.}$) and a Riemannian metric $g$ (from the real part of $\braket{.}{.}$). Then, the algebra of observables $\mathcal{B}_\mathbb{R}(\mathcal{H})$ can be recast as the space of functions of $\mathcal{C}^\infty ( P \mathcal{H}, \mathbb{R})$ which preserve both geometric structures, i.e. whose hamiltonian vector fields are also Killing vector fields.

Although it may look a bit abstract, this formulation frames quantum mechanics in very similar terms to classical mechanics, where the space of states is a symplectic manifold $(\mathcal{P},\omega)$, and the observables are functions of $\mathcal{C}^\infty(\mathcal{P},\mathbb{R})$. In this framework, both classical and quantum space of states are symplectic manifolds, but the quantum case bears the additional structure of a Riemannian manifold. One does not need to know the details of the geometrical formulation to understand the point we want to make, that is, classical mechanics can be seen as the particular case of quantum mechanics when the Riemannian structure is trivial, i.e. $g=0$!

This fact is of importance because the Riemannian metric gives precisely a measure of the uncertainty of observables. An observable $\hat A \in \mathcal{B}_\mathbb{R}(\mathcal{H})$, defines a function $A$ over $P\mathcal{H}$ through $A : \ket{\psi} \mapsto \ev{A}$, and thus a hamiltonian vector field $X_A$. One can prove that
\begin{equation}
\Delta \hat A = g(X_A,X_A).
\end{equation}
In the classical case ($g=0$), we have $\Delta \hat A = 0$  for any observable $\hat A$ and state $\ket{\psi}$. Classical mechanics is quantum mechanics without uncertainty.

One may wonder whether it was necessary to appeal to the abstract geometrical formulation to reach this conclusion. Indeed the result goes along very well with the intuitive idea that the quantum is fuzzy, while the classical is peaked. However, the geometrical formulation brings clarity and precision to the debate, and points towards a definite mathematical direction where to look for classicality inside the quantum realm.

The quest for "quasi-classical" states can now be reformulated in the following terms. Are there states $\ket{\psi}$ for which 
$\Delta \hat A = 0$ for any observable $\hat A$?

\passage{Eigenstates}

Start considering a single observable $\hat A$. What are the states that satisfy $\Delta \hat A = 0$? It is easily shown that they are all, and only, the eigenstates $\ket{a}$ of $\hat A$. Thus, an eigenstate shows some classical features, which is not a surprise after all: the eigenstate $\ket{a}$ of $\hat A$ is very peaked \textit{with respect to $\hat A$}. Similarly, $\ket{x}$ is classical in the sense that $\Delta \hat x = 0$, i.e. it is very peaked with respect to $\hat x$, which was indeed our first attempt to define "quasi-classical state" in section \ref{sec:dynamical characterisation}. So the last question of the previous paragraph, admits a direct answer: no. Because if $\Delta \hat A = 0$ for all $\hat A$, then $\ket{\psi}$ is an eigenstate of all $\hat A$ which is not possible.

Our expectations have to be qualified, and one can look instead for states which satisfy $\Delta \hat A = 0$ for \textit{some} observables $\hat A$, i.e. a common eigenstate of a subset $\mathcal{A} \subset \mathcal{B}_\mathbb{R}(\mathcal{H})$. Such an eigenstate can be said "quasi-classical" \textit{with respect to $\mathcal{A}$}. If $\mathcal{A}$ is commutative, then its common eigenstates form a basis of $\mathcal{H}$. Such are the eigenstates of a CSCO (complete set of commuting observables) which may be regarded in this respect as the most classical states of a given quantum system. However, if $\mathcal{A}$ is non-commutative, there will be generically no common eigenstates. The question is now shifted to the definition of "quasi-classical" states with respect to a non-commutative set of observables.

\passage{Squeezed coherent states}

Let's consider two non-commutative observables $\hat A$ and $\hat B$. Generically, they do not share any common eigenstates, so that we cannot have both $\Delta \hat A = 0 $ and $\Delta \hat B = 0 $. One has to find instead a fair trade-off between $\Delta \hat A$ and $\Delta \hat B$, so that they are both \textit{small}, although non zero. The trade-off is ruled by Heisenberg inequality which reads
\begin{equation}
\Delta \hat  A \Delta \hat B \geq \frac{1}{2} \left| \ev{[\hat A, \hat B]} \right|.
\end{equation}
One can show\footnote{See for instance \cite{hall2013} p. 244.} that this inequality is saturated precisely when $\ket{\psi}$ is an eigenstate either of $\hat A$, or of $\hat B$, or of 
\begin{equation}
\hat A + i \gamma \hat B
\end{equation} 
with $\gamma \in \mathbb{R}$. The meaning of this $\gamma$ is understood with the following corollary:
\begin{equation}
\gamma = \frac{\Delta \hat A}{\Delta \hat B}
\end{equation}
It ponders the respective weight of $\hat A$ and $\hat B$.
The eigenstates of $\hat A + i \gamma \hat B$ are called the \textit{$\gamma$-squeezed coherent states with respect to $\hat A$ and $\hat B$}.

\passage{Application to $\hat x$ and $\hat p$}

Let's apply the result to $\hat A = \hat x$ and $\hat B = \hat p$. 
Heisenberg inequality reads
\begin{equation}
\Delta \hat x \Delta \hat p \geq \frac \hbar 2.
\end{equation}
The normalised eigenstate of $\hat x + i \gamma \hat p$, with eigenvalue $z \in \mathbb{C}$, is 
\begin{equation}\label{eq:annihilation eigenstates}
\psi_{z, \gamma} (x) = \frac{e^{- \frac{(\Im z)^2}{2 \gamma \hbar}}}{\sqrt[4]{\pi \gamma \hbar}} e^{- \frac{(x - z)^2}{2 \gamma \hbar}}
\end{equation}
The normalisation is only possible for $\gamma > 0$. Thus, the squeezed coherent states (with respect to $\hat x$ and $\hat p$) form a $3$-dimensional submanifold of $P L^2(\mathbb{R})$, parametrised by $\gamma$ and $z$. 

The Schrödinger coherent states of equation \eqref{eq:schrodinger CS} are recovered by fixing $\gamma = \frac{1}{m \omega} $. More precisely, we have
\begin{equation}
\braket{x}{A e^{i \omega t}} =e^{i \frac{A^2}{2} \sin(2 \omega t)}  \psi_{z,\gamma}(x)
\end{equation}
with
\begin{align}
\gamma = \frac{1}{m \omega} && \text{and} &&z = \sqrt{\frac{2 \hbar}{m \omega}} A e^{i \omega t}.
\end{align}
We have found an equivalent definition of Schrödinger coherent states: they are the states which minimise $\Delta \hat x \Delta \hat p$, with equal weight\footnote{I.e. $\Delta \hat p = m \omega \Delta \hat x$. The constant $m \omega$ guarantees the homogeneity of the physical dimension.} for $\hat x$ and $\hat p$. 

\passage{Kinematics vs dynamics}

This new characterisation of coherent states differs from the previous one of Schrödinger in a central aspect: it only refers to the \textit{kinematics}, and not to the \textit{dynamics}. Indeed, the previous definitions were involving the specific form of the hamiltonian $\hat H$ of the harmonic oscillator, while now the definition only uses two observables $\hat x$ and $\hat p$ acting on the Hilbert space $L^2(\mathbb{R})$.

The move is noticeable because, for instance, the free particle and the harmonic oscillator have the same kinematics, and only differ by their dynamics. From this new kinematical characterisation, the previous coherent states of the harmonic oscillator could be equally called coherent states of the free particle, while the original dynamical characterisation doesn’t allow such a possibility.

To be clear, the family of coherent states, as a whole, can be fully characterised by kinematical considerations, but it will only exhibit nice dynamical properties in a particular case. Indeed, the family is stable under the harmonic oscillator evolution, whereas it is not with the free particle one. 

The transitional role from the dynamical to the kinematical perspective is played by the operator
\begin{equation}
\hat a \overset{\rm def}= \sqrt{\frac{m \omega}{2 \hbar}} \left( \hat x + \frac{i}{m \omega} \hat p \right).
\end{equation}
The task of minimising Heisenberg inequalities has been shown to reduce to that of finding the eigenstates of this operator, which are precisely the Schrödinger coherent states. This way, the operator $\hat a$ has arisen through purely kinematical considerations. However, the same operator plays an important role in the dynamics of the harmonic oscillator, where it is known as the \textit{annihilation operator}, for it acts destructively over the eigenstates of $\hat H$:
\begin{equation}
\hat a \, \ket{n} = \sqrt{n} \, \ket{n-1}.
\end{equation}
It is important to keep in mind this double role of $\hat a$ to understand better later generalisations of coherent states. 

\section{Optical coherence} \label{sec:optical coherence}

In the previous section, we have been looking for quasi-classical states and ended up with an abstract definition of coherent states as eigenstates of the so-called annihilation operator $\hat a$. This definition is the one used in many textbooks to define coherent states at first, but it is disappointing to see that it is often not motivated by physical considerations. We have shown how it could be motivated in a rather abstract way, from the geometrical formulation and the minimisation of Heisenberg inequalities. We are now going to show a more experimentally grounded way to introduce it, which is also the path that was followed historically. It is the way of Glauber when he revived coherent states, thirty years after Schrödinger, in the concrete context of quantum optics. Besides, we will understand why the adjective "coherent" can be preferred to "quasi-classical". The formulas of this section are taken from the book of recollections \cite{glauber2007}.

\passage{Quantum optics}

Quantum optics describes light using the theory of quantum electrodynamics (QED). The electromagnetic field is described by a state in a Hilbert space, and the observable quantities are described by the electric and magnetic hermitian operators, $\hat{\vec{ E}}(\vec r,t)$ and $\hat{\vec{ B}}(\vec r,t)$ (we use Heaviside-Lorentz units). 
In absence of any sources, the time evolution of states is driven by the hamiltonian
\begin{equation}
\hat H = \frac{1}{2} \int \dd \vec r \, (\hat {\vec E}^2 + \hat {\vec B}^2).
\end{equation} 
In the time gauge, these observables can in fact be derived from a vector potential $\hat{\vec{A}}$ such that
\begin{align} \label{eq:E=f(A)}
\hat{\vec{E}} = - \frac{1}{c} \pdv{\hat{\vec{A}}}{t} && \text{and} && \hat{\vec{B}} = \nabla \times \hat{\vec{A}}.
\end{align}
Assuming the field is confined within a cubic box of side $L$, the vector potential $\hat{\vec{A}}$ can be decomposed into a superposition of modes $k$ such that
\begin{multline} \label{eq:fourier(A)}
\hat{\vec{A}}(\vec r,t ) = c \sum_k \left( \frac{\hbar}{2 L^3 \omega_k} \right)^{1/2} ( \hat a_k \, \vec{e}_\lambda e^{i (\vec k \cdot \vec r - \omega_k t)} \\ + \hat a_k^\dagger \, \vec{e}_\lambda e^{ - i (\vec k \cdot \vec r - \omega_k t)} ),
\end{multline}
where the sum is made over an index $k$, used as a shorthand for $(\lambda, \vec k)$, where $\vec e_\lambda$ ($\lambda \in \{ 1,2 \}$) is the polarisation vector, perpendicular to $\vec k$, and $\vec k$ ranges over a discrete set of values permitted by the boundary conditions. Then $\hat a_k$ and $\hat a^\dagger_k$  are operators associated respectively to the positive and negative frequency part of $\hat{\vec{A}}$. They satisfy 
\begin{align}
[\hat{a}_k,\hat{a}_{k'}] =[\hat{a}^\dagger_k,\hat{a}^\dagger_{k'}] = 0 && \text{and} && [\hat{a}_k, \hat{a}^\dagger_{k'}] = \delta_{kk'}.
\end{align}
The hamiltonian can be rewritten as
\begin{equation}
\hat H =  \sum_k \hbar \omega_k (\hat{a}_k^\dagger \hat{a}_k + \frac{1}{2}).
\end{equation}
Thus, in absence of sources, the electromagnetic field is mathematically equivalent to an assembly of one-dimensional harmonic oscillators (one per mode $k$), so that $\hat a_k$ and $\hat a^\dagger_k$ are properly annihilation and creation operators. The basis of eigenstates of $\hat H$ is immediately deduced:
\begin{equation}\label{eq:stationary-free}
\bigotimes_k \ket{n_k}
\end{equation}
where $n_k$ is the number of photons in the mode $k$.

\passage{Interacting theory}

The basis of stationary states of the free theory, equation \eqref{eq:stationary-free}, is not the best suited for the description of states of light coming out of photon beams. Instead, the family of coherent states is much more convenient, and it appears naturally once one considers interactions.

So far, we have described the free theory of the electromagnetic field. But light is created by sources, like lamps or antennae, which consist of moving charges that excite the electromagnetic field. It is thus crucial to describe the interaction of light with charged matter, and notably to model the photon field radiated by a classical electric current. A classical current $\vec j (\vec r ,t)$ is assumed to interact with the vector potential through the following hamiltonian of interaction:
\begin{equation}\label{eq:H_I}
\hat{H}_I = \frac{1}{c} \int \vec j \cdot \hat{\vec A} \, \dd \vec r.
\end{equation}
Starting at initial time in the vaccum state $\ket{0}$, the field gets excited, and ends up at time $t$ in a state
\begin{equation} \label{eq:time evolution interaction}
\ket{t} = e^{i \phi(t)} e^{\frac{i}{\hbar} \int_0^t \hat{H}_I(t') \dd t'} \ket{0}
\end{equation}
where the phase $\phi(t)$ admits a definite expression, but irrelevant for our purposes. It can be rewritten
\begin{equation}
\ket{t} = e^{i \phi(t)} \bigotimes_k \ket{\alpha_k(t)},
\end{equation}
where $\ket{\alpha_k(t)}$ is the coherent state with
\begin{equation} \label{eq:alpha k t}
\alpha_k(t) = \frac{i}{\sqrt{2 L^3 \hbar \omega_k}} \int_0^t \vec j \cdot \vec e_\lambda e^{- i (\vec k \cdot \vec r + \omega_k t')} \dd \vec r \dd t'.
\end{equation}
This hamiltonian of interaction is a good model for most of the macroscopic sources where radiation is generated by a charged current $\vec j(\vec r,t)$ whose expression is known. In practice, lasers indeed produce coherent states of light, but incandescent bulbs do not, for they consist of many independent and chaotic sources which break the overall coherence.

Thus coherent states have appeared as the most natural states of the electromagnetic field when it is minimally coupled to a classical source. In addition to this special role in the production of light, coherent states exhibit major features from the point of view of its detection, which provides a clearer operational meaning to coherence, as we now explain.

\passage{Maximising interference}

To detect light, one uses a photon counter. Typically, a photon counter is sensible to the intensity of the electric field $E$, that we now assume to be only a scalar field, for simplicity.
If you assume that the electromagnetic field is in a state $\ket{\psi}$, then the intensity in $x = (r,t)$ (the spacetime point where the detector is) is on average
\begin{equation} \label{eq:I=E^2}
I(x)=\ev{\hat{E}^{(-)}(x)\hat{E}^{(+)}(x)}{\psi}
\end{equation}
where $\hat{E}^{(-)}$ is the negative frequency part of the electric field, conjugate to the positive frequency part $\hat{E}^{(+)}$, which is (from equations \eqref{eq:E=f(A)} and \eqref{eq:fourier(A)})
\begin{equation}\label{eq:E+}
\hat{E}^+(\vec r,t) = i \sum_k \left( \frac{\hbar \omega_k}{2L^3} \right)^{1/2} \hat{a}_k \, e^{i(\vec k \cdot \vec r - \omega_k t)}.
\end{equation}
Equation \eqref{eq:I=E^2} is a fancy way of writing that the energy of the electric field is $|E|^2$.

Let's define the first-order (or two-point) correlation function as
\begin{equation}\label{eq:def G1}
G(x_1,x_2) \overset{\rm def}= \ev{\hat{E}^{(-)}(x_1)\hat{E}^{(+)}(x_2)}{\psi}.
\end{equation}
In a double-slit experiment, the interference pattern observed on the screen is a measure of the intensity $I(x)=G(x,x)$ along the screen. In fact the electric field $E(x)$ on the screen is the linear superposition of the electric field $E(x_1)$ and $E(x_2)$ that was emitted by each of the two slits at spacetime points $x_1$ and $x_2$: 
\begin{equation}
\hat{E}(x) \propto \hat{E}(x_1) + \hat{E}(x_2).
\end{equation}
In this equation, $x$, $x_1$ and $x_2$ are related by the condition that spherical waves emitted at spacetime points $x_1$ and $x_2$ intersect in $x$.
As a consequence,
\begin{equation}
I(x) =  G(x_1,x_1) + G(x_2,x_2) + 2 \Re G(x_1,x_2).
\end{equation}
The two first terms are the independent contributions from each slit. The last term is responsible for the interference.
When $G(x_1,x_2)=0$, no fringes are observed. In fact, the visibility of the fringes is given by
\begin{equation}
v \overset{\text{def}}= \frac{I_{max} - I_{min}}{I_{max} + I_{min}} = \frac{2|G(x_1,x_2)|}{G(x_1,x_1) + G(x_2,x_2)}
\end{equation}
Now one can show the inequality:
\begin{equation}
|G(x_1,x_2)|^2 \leq G(x_1,x_1) G(x_2,x_2)
\end{equation}
so that, keeping $G(x_1,x_1)$ and $G(x_2,x_2)$ fixed, the maximum of interference is obtained for 
\begin{equation}
|G(x_1,x_2)| = \sqrt{G(x_1,x_1) G(x_2,x_2)}.
\end{equation}
When this condition is assumed to be valid for all $x_1$ and $x_2$, one can show that there exists a function $\mathcal{E}(x)$ so that
\begin{equation}
G(x_1,x_2) = \mathcal{E}^*(x_1) \mathcal{E}(x_2).
\end{equation}
$G(x_1,x_2)$ factorises, and the state $\ket{\psi}$ is said to be \textit{optically coherent}. This notion of coherence recovers one that already existed in classical electromagnetism, prior to quantum optics.

It is easy to see from the definition \eqref{eq:def G1} that a sufficient condition for the factorisability of $G(x_1,x_2)$ is that $\ket{\psi}$ is an eigenstate of $\hat{E}^{(+)}(x)$ for all $x$. From equation \eqref{eq:E+}, we see that it is equivalent to say that $\ket{\psi}$ is an eigenstate of $a_k$ for all $k$. And here we land on our feet! This is indeed the definition of coherent states that was given previously but applied to an assembly of independent harmonic oscillators. Here the definition is motivated on strong physical ground: the maximisation of the inference pattern or say differently the factorisation of the 2-point correlation function!

However, being a coherent state is only a sufficient, and not a necessary condition to be optically coherent, i.e. to factorise the 2-point correlation function. The coherent states can do much more: they factorise all of the higher-order correlation functions, which experimentally measures the coincidence rate between many detectors\footnote{However, it is unknown to the author whether a pure state that factorises all of the higher-order correlation functions is necessarily a coherent state of harmonic oscillator.}.

\passage{Coherence, Classicality, and Purity}

At this stage, the origin of the word "coherent" has been brought to light: the state $\ket{\psi}$ is such that the values of the field at different points of space-time "conspire" together to maximise the interference pattern. Meanwhile, we have lost sight of the sense in which they can be seen as "quasi-classical". Even worse, coherence and classicality may seem contradictory. Indeed, an example of coherent light is that produced by a laser, which is usually presented as a very quantum device, far from anything classical. On the contrary, ordinary light produced by incandescent bulbs, close to black-body radiation, is optically very incoherent, while it seems to be much more "classical" than lasers. Where is the catch?

The paradox arises from the confusion of two layers of "classicality". The first layer is classicality as the minimisation of the uncertainties of some non-commuting observables. Compared to the previous example of the harmonic oscillator, the vector potential $\hat A$ and the electric field $\hat E$ play now respectively the role of the position $\hat x$ and the momentum $\hat p$. Coherent states of light are quasi-classical in the sense that they minimise $\Delta \hat A$ and $\Delta \hat E$ together. 

The second layer of classicality is the difference between pure and mixed states. Classical physics is usually very noisy, that is very mixed, due to the difficulty to control interactions with the environment. For instance, the light of an incandescent bulb is a very mixed state (black-body radiation is maximally mixed), so that it is tempting to say that it is more classical with respect to a pure state, which is much more difficult to create in a lab.

The two layers, coherent/incoherent and pure/mixed, shall not be confused and are actually independent. Many pure states are incoherent, while some mixed states can be coherent. For instance, the state of an ideal laser is actually a mixed state, which reads
\begin{equation}
\rho = \frac{1}{2\pi} \int_0^{2 \pi} \dyad{|\alpha|e^{i \theta}} d\theta,
\end{equation}
although it is optically coherent (it factorises the $2$-point correlation function). This distinction is sometimes overlooked, especially in the context of quantum gravity, where one usually exclusively considers pure coherent states. This state of research may surprise as it is reasonable to believe that quantum states of space are horrendously hard to isolate. Considering mixed states instead may have some relevance in solving some of the hard problems in the field \cite{Amadei2019}. 

\section{Algebraic approach} \label{sec:algebraic approach}

\passage{Displacement operator}

\noindent
The vacuum state is the only coherent state which is also an eigenstate of of the harmonic oscillator: 
\begin{equation}
\ket{\alpha = 0} = \ket{n = 0}.
\end{equation}
For this reason, there is no ambiguity and one can write $\ket{0}$\footnote{Whereas it would be ambiguous to write for instance $\ket{1}$, since $\ket{\alpha = 1} = e^{-1/2} \sum_{n=0}^\infty \frac{1}{\sqrt{n!}} \ket{n} \neq \ket{n = 1}$.}. In the previous section, we have seen how the other coherent states are generated from the vacuum $\ket{0}$ by the unitary evolution of a simple hamiltonian of interaction $\hat H_I$. Equation \eqref{eq:time evolution interaction} can be rewritten
\begin{equation}
\ket{t} = e^{i \phi(t)} \prod_k \hat D_k(\alpha_k(t)) \ket{0}
\end{equation}
with $\alpha_k(t)$ given by equation \eqref{eq:alpha k t}, and $\hat D$ a unitary operator-valued function over $\mathbb{C}$, called the \textit{displacement operator}, and defined by 
\begin{equation}
\hat D(\alpha) \overset{\rm def}= e^{\alpha \hat a^\dagger - \alpha^* \hat a}.
\end{equation}
It is easy to check indeed that
\begin{equation}
\ket{\alpha} = \hat D(\alpha) \ket{0}.
\end{equation}

\passage{Heisenberg group}

The set of displacement operators almost form a group:
\begin{equation}\label{eq:D additivity}
 \hat D(\alpha + \beta) = e^{-i\Im (\alpha \beta^*)} \hat D(\alpha) \hat D(\beta)
\end{equation} 
In other words they form a group up to a phase. More precisely, they form a subset of a group, called the Heisenberg group. Let's see what it is.

The position and momentum operators $\hat x$ and $\hat p$ generate a Lie algebra called the \textit{Heisenberg algebra} (or also the \textit{Weyl algebra}). It is the smallest algebra generated by $\hat x$ and $\hat p$, with linear combination and Lie bracketing $i[.,.]$. It is a $3$-dimensional non-commutative real algebra, denoted $\mathfrak{h}$ and consisting of elements of the form
\begin{equation}
a \hat x + b \hat p + c \hat I \quad a,b,c \in \mathbb{R},
\end{equation}
where $\hat I$ is the identity operator.

Exponentiating the Heisenberg algebra gives a $3$-dimensional real Lie group, which is called, wisely, the \textit{Heisenberg group} (or also the \textit{Weyl group}), denoted $H_3$. It consists of elements of the form
\begin{equation}
e^{i(a \hat x+b \hat p+c \hat I)} \quad a,b,c \in \mathbb{R}.
\end{equation}
The displacement operators are just a subset of this group, such that
\begin{equation}
\hat D(\alpha) = e^{i \sqrt{2} \Im \alpha \hat x - i \sqrt{2} \Re \alpha \hat p }.
\end{equation}
Any element of $H_3$ can be written as a displacement operator times a phase. But since the global phase of states is irrelevant, the action of $H_3$ on $\ket{0}$ generates exactly the family of coherent states!

In fact, from equation \eqref{eq:D additivity}, it is easy to see that one can generate the whole family of coherent states starting from any $\ket{\alpha}$, and not only $\ket{0}$. One says that the action of the Heisenberg group is transitive: any two coherent states are related by a transformation of the Heisenberg group.  

\passage{Generalisation}

The previous analysis motivates a generalisation of coherent states for any Lie group $G$ acting over a Hilbert space $\mathcal{H}$, which was first proposed in parallel by Perelomov \cite{perelomov1972, perelomov1986} and Gilmore \cite{gilmore1974}.

Let $G$ be a Lie group, and $T$ a unitary irreducible representation (irrep) of $G$ over a Hilbert space $\mathcal{H}$. Choose $\ket{\psi_0} \in \mathcal{H}$, and denote $H$ the subgroup which stabilises $\ket{\psi_0}$ up to a phase, i.e.
\begin{equation}
 H \overset{\text{def}}= \left\{ g \in G \mid \exists \phi \in \mathbb{R}, \quad  T(g)\ket{\psi_0} = e^{i\phi} \ket{\psi_0} \right\}.
\end{equation}
The family of generalised coherent states is defined as the orbit of $\ket{\psi_0}$ under the action of the (left) quotient space $G/H$. More precisely, for each class $x \in G/H$, choose a representative $g(x) \in G$, and define the generalised coherent states as
\begin{equation}
\ket{x} = T(g(x)) \ket{\psi_0}.
\end{equation}
Thus, the generalisation of coherent states depends a priori on many choices: a group $G$, a unitary irrep $T$, a vector $\ket{\psi_0}$ and a set of representatives $g(x)$. Of course, when it is projected down to the projective Hilbert space $P\mathcal{H}$, the set of coherent states does not depend on the choice of the representatives $g(x)$. The choice of initial state $\ket{\psi_0}$ is a priori arbitrary, but can be motivated by another criterion, like being a ground state or minimising some uncertainty relations.

In the case of Schrödinger coherent states $\ket{\alpha}$, the group is $H_3$, with stabiliser $U(1)$, the initial state is the ground state $\ket{0}$ of the harmonic oscillator, and the representatives are the displacement operators $D(\alpha)$. We see again the interplay between the kinematical and dynamical sides of coherent states: the dynamics lies in the choice of the ground state $\ket{0}$, and the full set of $\ket{\alpha}$ is generated by the group $H_3$, which contains the kinematical aspects.
		
\passage{Bloch states}

When we apply the method to the basic Lie group $SU(2)$, we obtain what quantum opticians call the \textit{Bloch states}. $SU(2)$ is the exponentiation of the real Lie algebra $\mathfrak{su}(2)$, spanned by the (imaginary) Pauli matrices $(i \sigma_1, i\sigma_2, i\sigma_3)$. In other words, any $u \in SU(2)$ can be written as
\begin{equation}
u = e^{ i \vec \alpha \cdot \vec \sigma}, \quad \alpha \in \mathbb{R}^3.
\end{equation}
The unitary irreps of $SU(2)$ are Hilbert spaces $\mathcal{H}_j$, labelled by a spin $j \in \mathbb{N}/2$ and spanned by the magnetic basis $\ket{j,m}$, $m \in \{-j,...,j\}$, which diagonalises $J_3$ and $\vec J^2$ (in physicists notations, $J_i \overset{\rm def}= \frac{\sigma_i}{2}$), such as
\begin{equation}
\begin{array}{l}
J_3 \ket{j,m} = m \ket{j,m}, \\
\vec J^2 \ket{j,m} = j(j+1) \ket{j,m}.
\end{array}
\end{equation}
As an initial "vacuum" state we choose\footnote{The choice $\ket{j,j}$ is often made too.} $\ket{j,-j}$. Then, we can show that the stabiliser is $U(1)$, and we have the following diffeomorphism $SU(2)/U(1) \cong S^2$. The unit sphere $S^2$ can be parametrised by a complex number $\zeta \in \mathbb{C} $ (except for one point), by (the inverse of) the stereographical projection
\begin{equation}\label{eq:inverse stereo}
\vec n (\zeta) = \frac{1}{1+|\zeta|^2} \begin{pmatrix}
-\zeta - \zeta^* \\ i(\zeta^* - \zeta) \\ 1-|\zeta|^2
\end{pmatrix}.
\end{equation}
The representative $u \in SU(2)$ for each class $\vec n(\zeta) \in S^2$ is given by
\begin{equation}\label{eq:Perelomov map}
u(\zeta) \overset{\rm def}= \frac{1}{\sqrt{1+|\zeta|^2}} \begin{pmatrix}
1 & \zeta \\ -\zeta^* & 1
\end{pmatrix}.
\end{equation}
Finally we define the $SU(2)$ coherent states as
\begin{equation}\label{eq:SU(2) coherent states}
\ket{j,\vec n(\zeta)} \overset{\rm def}= u(\zeta) \ket{j,-j}.
\end{equation}
In terms of the magnetic basis, one can show that
\begin{equation}
\ket{j, \vec n(\zeta)} =  \frac{1}{(1+|\zeta|^2)^j} \sum_{m=-j}^j \binom{2j}{j+m}^{\frac 1 2} \zeta^{j+m} \ket{j,m}.
\end{equation}

Among the important properties that these states satisfy, we should note that the $SU(2)$ coherent states are eigenstates of $\vec n \cdot \vec J$
\begin{equation}
\vec n \cdot \vec J \ket{j,\vec n} = - j \ket{j,\vec n}.
\end{equation}
Also they saturate the following Heisenberg inequality
\begin{equation}
\Delta J_1  \Delta J_2 \geq \frac{1}{2} |\ev{J_3}|.
\end{equation}
Finally, they satisfy the following resolution of the identity
\begin{equation}
\frac{2j+1}{4 \pi} \int_{S^2} \dd \vec{n} \ket{j,\vec{n}} \bra{j,\vec{n}} = \mathbb{1},
\end{equation}
with $\dd \vec n$ being the usual measure on the unit sphere $S^2$.
The Bloch states have latter been used in quantum gravity as we shall see in section \eqref{sec:quantum gravity}.

\passage{Dynamical group}

From the perspective of experimentalists, the Lie group $G$ is not something abstract but something very concrete, for its action drives the unitary time evolution of states. In quantum optics, one deals typically with some effective model of perturbed hamiltonian:
\begin{equation}
\hat H = \hat H_0 + \hat H_{pert}.
\end{equation}
The initial state is chosen to be the ground state of $\hat H_0$, and the coherent states are generated through the time evolution induced by the perturbation $\hat H_{pert}$, which can be due to the coupling to some classical current as in equation \eqref{eq:H_I}.

In this context, the group $G$ is sometimes called a \textit{dynamical symmetry group}, so that for instance $H_3$ is said to be \textit{the} dynamical (symmetry) group of the harmonic oscillator \cite{gilmore1974, feng1990}. This naming is confusing because it conflicts both with the notion of "dynamical group", as defined by Souriau in \cite{Souriau1997}, and with the usual notion of "symmetry group" of a hamiltonian.

Usually, a classical physical system is given by a phase space $(\mathcal{P}, \omega)$ and a hamiltonian $H$. Souriau defines a dynamical group as any Lie group $G$ acting as a symplectomorphism (canonical transformation) over $\mathcal{P}$. Then one can consider the symmetry of the hamiltonian, i.e. the functions $C_i \in \mathcal{C}^\infty(\mathcal{P},\mathbb{R})$ such that 
\begin{equation}
\{ C_i, H \} = 0.
\end{equation}
They generate a Lie algebra of conserved quantities, which can be exponentiated into a Lie group, which is called the \textit{symmetry group} of the hamiltonian $H$. This group is acting over the phase space as symplectomorphism and for that reason, it is sometimes emphasised as the \textit{dynamical symmetry group} of $H$. In this sense, $H_3$ is \textit{not} the dynamical symmetry group of the harmonic oscillator!

This should not be too much of a surprise because, as we said earlier, the Schrödinger coherent states have little to do with the \textit{dynamics} of the harmonic oscillator. What might be regarded as dynamical in them is the initial choice of the ground state $\ket{0}$, but the Heisenberg group that further generates them is built by from a choice of coordinates $(x,p)$ over the phase space, that is, independently of the specific form of the hamiltonian of the harmonic oscillator.

\section{Geometric approach} \label{sec:geometric approach}

In the two previous sections, we have seen how some initial quantum optical work by Glauber in the 1960s \cite{glauber2007}, has lead in the 1970s, to a generalisation of coherent states, by Perelomov \cite{perelomov1972} and Gilmore \cite{gilmore1974}, using Lie groups. In this section, we explore a second and independent path of generalisation in more geometrical terms, which was proposed in the 1990s by Hall \cite{hall1994, hall2002}, based on some earlier works by Segal \cite{Segal1960} and Bargmann \cite{Bargmann1961} in the 1960s. Both approaches have their relevance for quantum gravity, as we shall see in section \ref{sec:quantum gravity}.

\passage{Phase space vs Hilbert space}

The classical phase space of the harmonic oscillator is $T^*\mathbb{R}$, endowed with the usual symplectic structure given by the determinant. The quantum analogue is the Hilbert space $L^2(\mathbb{R})$, with the usual scalar product. The family of coherent states constitute a $2$-dimensional submanifold of $L^2(\mathbb{R})$, parametrised by amplitude and time $(A, t)$, or the complex number $z = A e^{i \omega t}$. It could be as well-parametrised by position $x = A \cos \omega t$ and momentum $p = - A \omega \sin \omega t$, which exhibit an explicit diffeomorphism between the classical phase space and the family of coherent states. Any point in phase space determines uniquely a coherent state, and conversely. This fact is not a coincidence, but rather a crucial aspect of coherent states that shed further light on their classicality. 

It is a general feature of coherent states that they define a natural injection of the classical phase space $\mathcal{P}$ into the Hilbert space $\mathcal{H}$. A priori, there are many possible such injections, but coherent states provide a natural one. To work this out, we shall first see how one can build a Hilbert space $\mathcal{H}$ from a phase space $\mathcal{P}$. Well, this is the whole point of quantization, and so one should know that the subject is not easy. Nevertheless \textit{geometric quantization}\footnote{Not to be confused with the previously discussed geometrical formulation of quantum mechanics \cite{Schilling1996}.} is a prototypical such method, rather technical, but we can skip the details and keep the general idea.

\passage{Geometric quantisation}

Start with a configuration space $\mathcal{M}$ and build the phase space $\mathcal{P} = T^*\mathcal{M}$. There are actually many ways in which $\mathcal{M}$ can be seen as a subspace of $T^*\mathcal{M}$. Each way consists in choosing what is called a \textit{polarisation}. Then one can build a complex line bundle over $T^*\mathcal{M}$, denoted $L$. Locally, we have $L \approx T^*\mathcal{M} \times \mathbb{C}$. The \textit{prequantum Hilbert space}, $\mathcal{PH}$, is the space of equivalence classes of square-integrable sections of $L$, where two sections are said equivalent when they are equal almost everywhere. Roughly, $\mathcal{PH} \approx L^2(T^*\mathcal{M})$. It is much too big to be a good quantum Hilbert space. So a choice of polarisation enables to select a subspace of $\mathcal{PH}$ and to build the good quantum Hilbert space $\mathcal{H} \cong  L^2(\mathcal{M})$. This construction really enable to see the Hilbert space $\mathcal{H}$ as a subspace of $L^2(T^*\mathcal{M})$. Thus a state $\ket{\psi} \in \mathcal{H}$ can be seen as a complex function $\psi$ over $T^*\mathcal{M}$.

Then, for each phase space point $z \in T^*\mathcal{M}$, we define the coherent state $\ket{z}$ as the unique state in $\mathcal{H}$ such that 
\begin{equation}\label{eq:CS phase space}
\forall \ket{\psi} \in \mathcal{H}, \quad \braket{z}{\psi} = \psi(z).
\end{equation}
This definition is both elegant and confusing. Elegant, because the equation is very simple, confusing, because it is too simple. On the LHS, we have a scalar product between two states in $\mathcal{H}$, while on the RHS we have the evaluation of a function $\psi$ in one point $z$ of the phase space $T^*\mathcal{M}$. A practical example should clarify the matter. 

\passage{Segal-Bargmann transform}

The classical phase space of the harmonic oscillator is $T^* \mathbb{R} \cong \mathbb{C}$, so that the prequantum Hilbert space is roughly $L^2(\mathbb{C})$. By using the Kähler polarisation, the quantum Hilbert space finally obtained is the Segal-Bargmann space\footnote{Also called the Fock-Bargmann space in \cite{gazeau2009}.} \cite{hall1994}, denoted $\mathcal{SB}$. It is made of functions over $\mathbb{C}$ which are both holomorphic and square-integrable, with the (gaussian) scalar product
\begin{equation}
(f_1,f_2) = \int_\mathbb{C} \bar{f_1}(z) f_2(z) \frac{i}{2\pi \hbar} e^{- |z|^2 /\hbar} dz \wedge d\bar{z} 
\end{equation}
$\mathcal{SB}$ is isomorphic to $L^2(\mathbb{R})$, which is not a big surprise after all, since all Hilbert spaces of the same dimension are isomorphic. What is more interesting is that there is actually an isometry between them two, given by
\begin{equation}\label{eq:isomorphism SB L2}
\tilde \phi(z) = \int_\mathbb{R} K(z,x) \phi(x) dx
\end{equation}
with the kernel
\begin{equation}
K(z,x) = \sqrt[4]{\frac{m \omega}{ \pi \hbar }} e^{\frac{1}{\hbar} \left( \frac{z^2}{2} - ( \sqrt{\frac{m\omega}{2}}x - z)^2 \right)}
\end{equation}
This isomorphism is called the Segal-Bargmann transform. Its inverse is given by
\begin{equation}
\phi(x) = \int_\mathbb{C} \overline{K(z,x)} \tilde \phi(z)  \frac{i}{2\pi \hbar} e^{- |z|^2 /\hbar} dz \wedge d\bar{z} 
\end{equation}
Equation \eqref{eq:isomorphism SB L2} is in fact a concrete instantiation of equation \eqref{eq:CS phase space}, so that the kernel $K$ is actually a coherent state! More precisely, it matches the expression of equation \eqref{eq:annihilation eigenstates}, provided a rescaling of $z$, and up to a phase and a normalisation factor:
\begin{equation}
K \left( z, x \right) =  e^{  \frac{i}{\hbar} \Re z \Im z} e^{\frac{1}{ \hbar} |z|^2}  \psi_{\sqrt{\frac{2}{m \omega}} z,\frac{1}{m \omega}} (x).
\end{equation}
For this reason the Segal-Bargmann transform is also called the coherent-state transform.

In the case of Schrödinger coherent states, the algebraic approach (with the Heisenberg group $H_3$) generates the same coherent states as the geometric approach (with the phase space $T^*\mathbb{R}$). Both approaches are secretly linked by the fact that the Heisenberg group $H_3$ is naturally obtained by exponentiating the quantised coordinates $\hat x$ and $\hat p$ of the phase space $T^*\mathbb{R}$. However, the two methods do not always match. In the case of the phase space $S^2$, the exponentiation of the quantised coordinates $\sigma_1, \sigma_2, \sigma_3$, generates $SU(2)$. However the coherent states obtained from geometric quantisation of $S^2$ \cite{hall2002a} are different from the $SU(2)$ coherent states of equation \eqref{eq:SU(2) coherent states}.

\passage{Resolution of the identity}

The fact that the coherent-state transform is an isometry is equivalent to the following \textit{resolution of identity} for the coherent states
\begin{equation}
\mathbb{1} = \frac{1}{\pi} \int_\mathbb{C} \dyad{\alpha} \, \dd \Re \alpha \, \dd \Im \alpha.
\end{equation}
This equation should be understood in the sense of weak convergence, that is, for any two given states $\ket{\phi_1}$ and $\ket{\phi_2}$,
\begin{equation}
\braket{\phi_1}{\phi_2} = \frac{1}{\pi} \int_\mathbb{C} \braket{\phi_1}{\alpha} \braket{\alpha}{\phi_2} \, \dd \Re \alpha \, \dd \Im \alpha.
\end{equation}
This resolution of the identity is similar to the more familiar one of any orthonormal basis of $\mathcal{H}$, like
\begin{equation}
\mathbb{1} = \sum_n \dyad{n},
\end{equation}
up to the crucial difference that the coherent states are parametrised by a \textit{continuous} parameter $\alpha \mapsto \ket{\alpha}$.

The (strong) continuity of the map $\alpha \mapsto \ket{\alpha}$ together with the (weak) resolution of identity are so important that they are often regarded as \textit{the} two properties that coherent states should have to deserve such a designation. It is a bit surprising because it seems too generic\footnote{In the same manner as the condition of being an orthonormal basis is far from being a sufficient property to define the $\ket{n}$ basis.}, but this is the point of view defended for instance by Klauder in his collection of papers \cite{Klauder1985}.

\passage{Heat kernel}

When the geometric quantisation is performed using the Khäler polarisation, the coherent states finally obtained by equation \eqref{eq:CS phase space} can be expressed in terms of the more baroque notion of the \textit{heat kernel} \cite{hall2002}. We explain it below for it has played a role in quantum gravity as we will see in the next section.

The heat kernel $\rho_t(x)$ is the solution of the heat equation
\begin{equation}
\frac{d\rho}{dt} = \partial_x^2 \rho
\end{equation}
that satisfies $\rho_0(x) = \delta(x)$.
Its explicit expression is\footnote{The heat equation is just Schrödinger equation with complex time, so that the solution can be easily recovered from \eqref{eq:free particle delta}. However, the solutions in \eqref{eq:free particle delta} were discarded as coherent states. It seems to be a pure coincidence that the same equation with complex time now gives proper coherent states.}
\begin{equation}
\rho_t(x) = \frac{1}{\sqrt{4\pi t}} e^{- \frac{x^2}{4t}}
\end{equation}
The kernel $K$ can be rewritten in term of the \textit{heat kernel} $\rho_t(x)$ such that
\begin{equation}\label{eq:CS heat kernel}
K \left(z\sqrt{\hbar} ,x \sqrt{\frac{\hbar}{2 m \omega}}\right) = \sqrt[4]{\frac{2 m \omega }{ \hbar}} \frac{\rho_{\frac 1 2}(x-z)}{\sqrt{\rho_{\frac 1 2}(x)}}.
\end{equation}

This observation has suggested a construction of coherent states when the configuration space is a connected compact Lie group $G$, instead of $\mathbb{R}$ \cite{hall1994}. The heat equation over $G$ reads
\begin{equation}
\frac{d\rho}{dt} =  \Delta \rho,
\end{equation}
where $\Delta$ is the Casimir operator, and $\rho$ a function of $t \in \mathbb{R}$ and $g \in G$. It can be shown the existence of a smooth and strictly positive solution, called the \textit{heat kernel}, which is a delta over the identity at $t=0$. It admits the following expansion
\begin{equation}
\rho_t(g) = \sum_\pi \dim \pi \, e^{- \lambda_\pi t } \, \chi^\pi(g),
\end{equation}
where the sum is taken over all irreps $\pi$ (up to unitary equivalence), and $\lambda_\pi$ is the characteristic Casimir (non-negative) number of the representation, and $\chi^\pi$ is the character. For instance, in the case of $SU(2)$ one gets
\begin{equation}\label{eq:heat kernel SU(2)}
\rho^o_t(g) = \sum_{j\in \mathbb{N}/2} (2j+1) \, e^{-j(j+1) t} \, \Tr D^j(g),
\end{equation}
with $D^j(g)$ the Wigner matrix of $g$ in the spin-$j$ irrep.
Then Hall defines the \textit{complexification} $G_\mathbb{C}$ of the Lie group $G$. For instance, for $SU(2)$ it is $SL_2(\mathbb{C})$. From this he shows that there is a unique analytic continuation of the heat kernel $\rho_t$ from $G$ to $G_\mathbb{C}$. In analogy\footnote{The square-root in equation \eqref{eq:CS heat kernel} is only a normalisation factor, of which one can get rid of, provided a good choice of measure in the definition of the scalar product.} with equation \eqref{eq:CS heat kernel}, the coherent states are defined as the functions over $G$, indexed by $x \in G_\mathbb{C}$, as
\begin{equation}\label{eq:CS heat kernel general}
K_{x,t}(g) \overset{\rm def}= \rho_t(x^{-1}g), \quad \text{with} \, g \in G.
\end{equation}

In \cite{hall2002}, it is shown how the heat kernel construction with a group $G$ matches the geometric quantisation approach over the phase space $T^*G$. The latter approach shows how the coherent states provide a natural embedding of the phase space within the corresponding Hilbert space, and in this sense, it points towards their quasi-classical properties. The heat kernel approach presents the advantage of offering more analytical formulas like \eqref{eq:CS heat kernel general}, compared to \eqref{eq:CS phase space}, but it is then harder to see how the quasi-classical properties can arise.

\section{Quantum Gravity} \label{sec:quantum gravity}

Quantum gravity is still at a very speculative stage compared to quantum optics, but the two fields of research can speak to one another. Indeed, the experimental control of the latter has motivated many theoretical developments, like the coherent states, which can now be reinvested into the former. At least, having in mind the concrete set-up of quantum optics may help theoreticians of quantum gravity to keep their feet on the ground, so to speak.

Schrödinger had introduced the coherent states as an attempt to recover the macroscopic physics from quantum mechanics. This motivation was somehow lost in quantum optics, which rather focused on interference patterns and addressed different kind of questions, closely related to technological achievements, such as lasers. In quantum gravity, the emphasis is put back on the classical properties of coherent states. This time, the coherent states are states of the gravitational field, whereas quantum optics deals with the electromagnetic field. The question here is to understand how general relativity can be recovered, in some limit, from quantum gravity, i.e. how a classical geometry of space-time can arise from quantum states of the gravitational field. Thus, coherent states are used as a tool to check the consistency of the theory in an experimentally well-tested macroscopic regime.

Taking the perspective of the kinematical characterisation described in section \ref{sec:kinematical characterisation}, one wonders which observables are to be chosen for the coherent states to be peaked upon a classical configuration of space-time. In quantum optics, we have seen that the relevant observables were the vector potential $\hat A$ and the electric field $\hat E$. General relativity can be formulated in close analogy to electromagnetism and this suggests to take as observables the so-called holonomy and flux, which carry the geometrical meaning of curvature and area.

Although any theory of quantum gravity is likely to make use of coherent states, we focus in this section on the Loop approach to Quantum Gravity (LQG). Both the algebraic (Lie group) and the geometric (heat kernel) approaches to coherent states have been applied in LQG. Historically, first has come the heat kernel method, starting with an extension of Hall's construction to diffeomorphism invariant gauge theories \cite{ashtekar1996e}. Bloch states were later used in \cite{livine2007b} and helped in providing a semi-classical picture of the geometry of space. But let's start recalling briefly the essential mathematical structure of LQG.

\passage{LQG Hilbert space}

\begin{figure}
\centering
\includegraphics[scale=1]{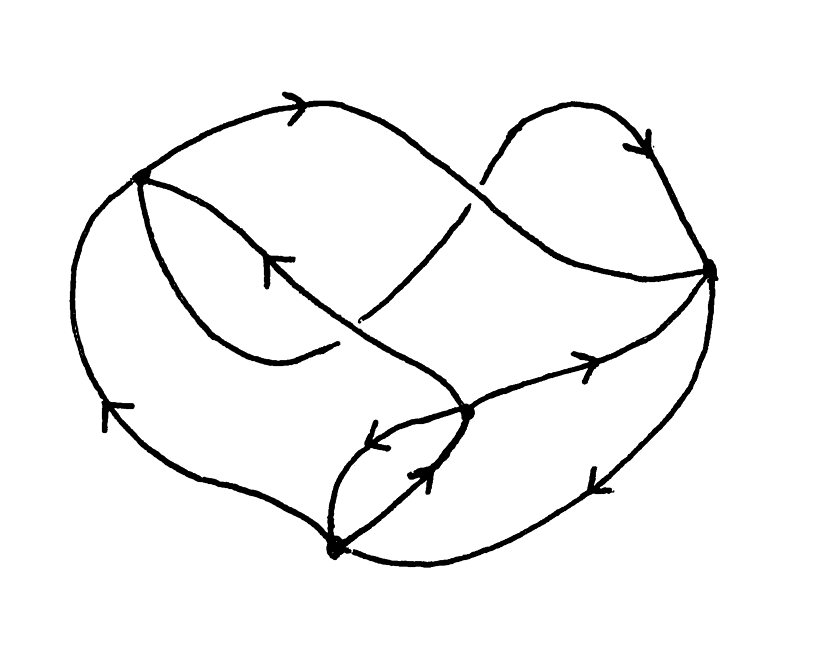}
\caption{Example of a finite directed graphs $\Gamma$, with $8$ links and $4$ nodes.}
\label{fig:graph}
\end{figure}

In covariant LQG, space is described by quantum states that form the following Hilbert space
\begin{equation}
\mathcal{H}_{LQG} = \bigoplus_\Gamma L^2(SU(2)^L)/\mathcal{G}_\Gamma,
\end{equation}
where the direct sum is made over all the different abstract finite directed graphs $\Gamma$, with different number of links $L$ and nodes $N$. An example of such a graph is given in figure \ref{fig:graph}. $L^2(SU(2)^L)$ is the Hilbert space onsisting of the square-integrable complex functions $\Psi(g_{l_1},...,g_{l_L})$ over $SU(2)^L$. It is quotiented by the so-called Gauss constraint $\mathcal{G}_\Gamma$, which amounts to imposing the following invariance property, given any element $u_n \in SU(2)$ per node $n$ of $\Gamma$,
\begin{equation}\label{eq:Gauss constraint functions}
\Psi(g_{l_1},...,g_{l_L}) =  \Psi(u_{t_1}g_{l_1}u^{-1}_{s_1},...,u_{t_L} g_{l_L}u^{-1}_{s_L}),
\end{equation}
with $t_i$ and $s_i$ being respectivelly the target and the source of the link $i$.

This description of quantum states of space might seem very abstract and one needs to draw a connection to some intuitive understanding in terms of geometric quantities. To make it short, the quantum space over a single link, $L^2(SU(2))$, is the quantisation of the phase space $T^*SU(2)$. Actually, since $SU(2)$ is a Lie group, its cotangent bundle can be trivialised as $T^*SU(2) \cong SU(2) \times \mathfrak{su}(2)$, and the coordinates $(h,E)$ are respectively called \textit{holonomy} and \textit{flux}. These quantities appear in general relativity, where they can be roughly understood as the curvature and the area of pieces of space. Subtlety left aside, this description of phase space suggests that it is possible to apply directly the heat kernel method described in section \ref{sec:geometric approach}, with the Lie group $SU(2)$! Thus we expect the coherent states of LQG to be indexed by elements of the complexification of $SU(2)$, that is $SL_2(\mathbb{C})$.

\passage{Coherent spin-networks}

The rigorous construction of coherent states for LQG was done in \cite{thiemann2001, thiemann2001b, thiemann2001c, thiemann2001a}, where they are called "complexifier coherent states", but we prefer to call them \textit{coherent spin-network} states as in \cite{bianchi2010c, bianchi2010d}. They are not properly coherent states for the full theory, but only for its truncation on a fixed graph $\Gamma$. They are denoted $\Psi_{(\Gamma,H_l,\tau_l)}$, parametrised by one element $H_l \in SL_2(\mathbb{C})$ and one positive number $\tau_l \in \mathbb{R}^+$ per link $l$ of $\Gamma$. To each link $l$ is associated the coherent state $\rho^o_{\tau_l} (g_l \, H_l^{-1})$, with $\rho^o_\tau(g)$ the analytic continuation  of \eqref{eq:heat kernel SU(2)} to $SL_2(\mathbb{C})$. All these terms are then multiplied together, and finally projected it down to $\mathcal{H}_{LQG}$ by an $SU(2)$ integration over each node to impose the Gauss constraint. All this fits in the following formula:
\begin{multline}
\Psi_{(\Gamma, H_l,\tau_l)}(g_l) \\ \overset{\rm def}= \int_{SU(2)^N} \left( \prod_l \rho^o_{\tau_l} (u_{t_l}\, g_l \, u^{-1}_{s_l} \, H_l^{-1}) \right) [\dd u_n] .
\end{multline}

The formal jungle of techniques that helps to define these coherent states in \cite{thiemann2001, thiemann2001b, thiemann2001c, thiemann2001a} should not make us forget about the underlying physical intuition that originally motivates them: they constitute a natural embedding of the classical phase space into the quantum Hilbert space. The embedding is parametrised by $H_l \in SL_2(\mathbb{C}) \cong T^*SU(2)$ which admits two possible semi-classical interpretations.

\passage{Holonomy and flux}

The first and original interpretation is based on the polar decomposition of $H$ as
\begin{equation}
H = h \, e^{i \frac{E}{8 \pi G \hbar \gamma} \tau},
\end{equation}
with $h \in SU(2)$ and $E \in \mathfrak{su}(2)$ being respectively the holonomy and the flux over a link, while $\gamma \in \mathbb{R}$ is a constant, called the Immirzi parameter, and $G$ and $\hbar$ respectively Newton's and (reduced) Planck's constant.
They are peaked on the classical holonomy-flux configuration $(h_l,E_l)$ on each link $l$ \cite{thiemann2001b, sahlmann2001}.

\passage{Twisted geometry}

The second interpretation is based instead on the decomposition
\begin{equation}
H = u(\zeta_s) \, e^{-i z \sigma_3} \, u(\zeta_t)
\end{equation}
with $\zeta_s, \zeta_t \in \mathbb{C}$, and $u(\zeta) \in SU(2)$ given by equation \eqref{eq:Perelomov map}, and 
\begin{equation}
z = \xi + i \,  a \, \tau
\end{equation}
with $\xi, a \in \mathbb{R}$.

With this parametrisation, the state is peaked on a discrete kind of geometry, called \textit{twisted geometry} \cite{freidel2010b}. The phase space $T^*SU(2)$ on each link is parametrised by $(\vec n(\zeta_s), \vec n (\zeta_t), \xi, a)$, with the map $\vec n (\zeta)$ given by equation \eqref{eq:inverse stereo}. Thus each node is surrounded by a set of unit vectors $\pm \vec n_i$, one per attached link, with the sign $\sigma = \pm$ chosen depending on whether the link is ingoing or outgoing. At the semi-classical level, the Gauss constraint imposes the closure of the vectors $\vec n_i$ surrounding a node, such as
\begin{equation}\label{eq:closure}
\sum_{l\in n} \sigma_l \, a_l \, \vec n_l = 0
\end{equation}
where the sum is done on all the links $l$ surrounding a node $n$.
It is a theorem by Minkowski that there is a unique convex polyhedron such that its faces have unit exterior-pointing normals $\sigma_l \, \vec n_l$ and areas $a_l$ satisfying the closure condition \eqref{eq:closure} \cite{minkowski1897, bianchi2011}. 
So each node of the coherent state can really be seen, in the semi-classical picture of twisted geometry, as a polyhedron.

Two neighbouring polyhedra are glued along a link, that is a face of same area, although the shape of the face may not match (reason for which the geometry is said to be \textit{twisted}).
Finally, the number $\xi$ can be used to encode the extrinsic curvature on the common face, when the polyhedra are embedded into $4$-dimensional spacetime\footnote{The issue is delicate, see \cite{anza2015}.}.

\passage{Coherent intertwiners}

The previous parametrisation is also interesting for it connects with the algebraic approach developed in section \ref{sec:algebraic approach}. In quantum gravity, the Lie group that generates the coherent states is $SU(2)$. It is not too much of a surprise since $SU(2)$ plays a central role in the description of the geometry of space. Thus, one recovers the so-called Bloch states $\ket{j, \vec n}$ (see equation \eqref{eq:SU(2) coherent states}), which are very familiar to quantum optics for very different reasons.

The Hilbert space $\mathcal{H}_{LQG}$ can be seen as a subspace of some direct sum of the Hilbert spaces $\mathcal{H}_j$ in which the Bloch states live. This suggests a second way to build coherent states for quantum gravity. Given a node $n$ of $\Gamma$, the idea is to select one Bloch state $\ket{j_l, \vec n_l}$ per link $l$ attached to $n$, to tensor them all and to project them down by imposing the Gauss constraint. One obtains the so-called \textit{coherent intertwiners} $\ket{\{ j , \vec n \}}_n$ \cite{livine2007b}: 
\begin{equation}
\ket{\{ j , \vec n \}}_n \overset{\rm def}= \int_{SU(2)} \left( \bigotimes_{l \in n} D^{j_l}(g) \ket{j_l, \vec n_l} \right) \dd g.
\end{equation}
Over a graph $\Gamma$, the tensor product of coherent intertwiners (one per node $n$ of the graph),
\begin{equation}
\bigotimes_{n \in \Gamma} \ket{\{ j , \vec n \}}_n,
\end{equation}
can be seen as a state 
\begin{equation}
\Psi_{\Gamma,j_l,\vec n_l} \in L^2(SU(2)^L)/ \mathcal{G}_\Gamma.
\end{equation}
The semi-classical interpretation of $ \Psi_{\Gamma,j_l,\vec n_l} $ is the juxtaposition of all polyhedra, i.e. without the extrinsic angle that glues them together. For this reason they are called \textit{intrinsic coherent states} by Rovelli \cite{rovelli2014a}, as opposed to the \textit{extrinsic coherent states} $\Psi_{(\Gamma, H_l,\tau_l)} $. For large values of $a$, the two are related by
\begin{multline}
\Psi_{(\Gamma, H_l,\tau_l)}(g_l) \\ \sim \sum_{j_l} \left( \prod_l (2j_l+1) e^{- \tau_l (j_l - \frac{a_l}{2} + \frac{1}{2})^2} e^{- i \gamma \theta_l j_l}\right) \\ \times \Psi_{\Gamma,j_l,\vec n_l} (g_l),
\end{multline}
(see \cite{bianchi2010c} for details).

\bigskip
We have given a quick overview of the two main definitions of coherent states in the context of quantum gravity. The main point was to show that both lines of investigation exposed previously, the algebraic and the geometric approaches, have spawned the field of quantum gravity. There are also other proposals under investigation like the $U(N)$ coherent states \cite{freidel2011}, the $SO^*(2N)$ coherent states \cite{Girelli2017}, or the coherent intertwiners of \cite{freidel2014}, but we postpone their review to future work.

\section{Conclusion}

The main results of this paper are the followings: 
\begin{itemize}
	\item We formalised in precise terms the hand-waving definition of the seminal paper of Schrödinger.
	\item We investigated what it would mean to be a coherent state of the free particle.
	\item We proposed the kinematical/dynamical dichotomy as a relevant conceptual shift towards the generalisation of the initial Schrödinger's idea.
	\item We noticed and resolved the paradox that the coherent states of light may appear as very non-classical states.
	\item We noticed and clarified a confusion in the literature in the use of the term of "dynamical symmetry group". 
	\item We summarised the genealogy of coherent states from Schrödinger to quantum gravity, enlighting the existence of two main parallel lines of thoughts, that were dubbed the algebraic and the geometric approaches.
\end{itemize}

Despite the broad range of our exploration, we have only been scratching the surface of the general subject of coherent states, which is now branching in many directions (see \cite{gazeau2009} or \cite{perelomov1986} for more exhaustive treatment, although quantum gravity is absent in them). Nevertheless, we have gathered in few pages, the core ideas of coherent states, as they have emerged in the head of Schrödinger, later fuelled by quantum optics, and applied to quantum gravity. These three steps of developments are, to some extent, independent, but we have shown how they mingle together to form a beautiful and consistent landscape.

\section*{Acknowledgements}

I would like to thank Alexander Thomas, Andrea Di Biagio, Andrea Calcinari, Carlo Rovelli, Emanuele Polino, Farshid Soltani, Federico Zalamea, John Schliemann, Klaus M\o lmer, Lautaro Amadei, Pietro Dona and Simone Speziale, for their comments, suggestions, encouragements and proofreadings.

\bibliographystyle{unsrtnat}
\bibliography{coherent}

\end{document}

%% file: config.tex
\pdfoutput=1

\usepackage{graphicx}
\usepackage{amsmath,amssymb,amsfonts}
\usepackage{hyperref}

\usepackage{physics} 
\usepackage[percent]{overpic} 
\usepackage{bbold} 

\usepackage[utf8]{inputenc} 
\usepackage[T1]{fontenc} 
\usepackage[english]{babel} 

\usepackage[numbers,sort&compress]{natbib}

\usepackage[autostyle, english = british]{csquotes}
\MakeOuterQuote{"}

\newenvironment{proof}
{\scriptsize
\textsc{Proof.}}
{$\Box$}

\newcommand{\passage}[1]{\bigskip
                         {\bf #1.}
                         \nopagebreak}